\documentclass[conference]{IEEEtran}
\IEEEoverridecommandlockouts
\usepackage{amsmath,url}

\usepackage{cite}
\usepackage{amsmath,amssymb,amsfonts, bm}

\usepackage{enumerate}
\usepackage[linesnumbered,ruled,vlined]{algorithm2e}
\let\oldnl\nl% Store \nl in \oldnl
\newcommand{\nonl}{\renewcommand{\nl}{\let\nl\oldnl}}% Remove line number for one line

\usepackage[]{graphicx}
\graphicspath{{Figures/}}
\usepackage{support-caption}
\usepackage{subcaption}
\usepackage{lipsum}
\usepackage{balance}
\usepackage{textcomp}

\usepackage{multirow}
\usepackage{color}
\usepackage{listings}
\usepackage{courier}
\usepackage{float}
\usepackage{verbatim}
\usepackage{array}

\usepackage{fancyhdr}
\usepackage{url}
\usepackage{hyperref}

\usepackage{booktabs,siunitx,rotating}

\definecolor{CustomColor}{RGB}{255, 255,255}

\lstdefinelanguage{turtle}
{
    columns=fullflexible,
    keywordstyle=\color{red},
    morekeywords={@prefix,@base,@forSome,@forAll,@keywords},
    morecomment=[l]{\#},
    tabsize=4, 
    alsoletter={-?}, % allowed in names
    morecomment=[s][\color{blue}]{<}{>},
    basicstyle=\ttfamily\color{black}, 
    morestring=[b][\color{black}]\",    
    backgroundcolor=\color{CustomColor}
}
\lstdefinestyle{turtle}{%
    morekeywords={a, @prefix},
    morecomment=[s][\rmfamily]{<}{>},
    morecomment=[s][\itshape]{"}{"},
}

\def\BibTeX{{\rm B\kern-.05em{\sc i\kern-.025em b}\kern-.08em
    T\kern-.1667em\lower.7ex\hbox{E}\kern-.125emX}}

\usepackage{nomencl}
\makenomenclature

\begin{document}

%=== Input Title here ================================

\title{Ontology Modeling for Decentralized\\ Household Energy Systems}

%===================================================

\author{
Jiantao~Wu$^{1,2}$,
Fabrizio Orlandi$^{1,3}$,
Tarek AlSkaif$^{4}$,
Declan O'Sullivan$^{1,3}$,
and~Soumyabrata~Dev$^{1,2}$\\
$^{1}$~ADAPT SFI Research Centre, Dublin, Ireland \\
$^{2}$~School of Computer Science, University College Dublin, Ireland\\
$^{3}$~School of Computer Science and Statistics, Trinity College Dublin, Ireland\\
$^{4}$~Information Technology Group, Wageningen University and Research, The Netherlands\\
\thanks{This research was partially funded by the EU H2020 research and innovation programme under the Marie Skłodowska-Curie Grant Agreement No.~713567 at the ADAPT SFI Research Centre at Trinity College Dublin. The ADAPT Centre for Digital Content Technology is funded under the SFI Research Centres Programme (Grant 13/RC/2106) and is co-funded under the European Regional Development Fund.}
\thanks{Send correspondence to S.\ Dev, email: soumyabrata.dev@ucd.ie}% <-this % stops a space 
}

% \author{
% \IEEEauthorblockN{Jiantao~Wu}
% \IEEEauthorblockA{\textit{ADAPT SFI Research Centre, Dublin} \\
% \textit{University College Dublin (UCD)}\\
% Dublin, Ireland\\
% mayank.jain1@ucdconnect.ie}
% \and
% \IEEEauthorblockN{Fabrizio Orlandi}
% \IEEEauthorblockA{\textit{ADAPT SFI Research Centre} \\
% \textit{Trinity College Dublin}\\
% Dublin, Ireland\\
% tarek.alskaif@wur.nl}
% \and
% \IEEEauthorblockN{Tarek~AlSkaif}
% \IEEEauthorblockA{\textit{Information Technology Group} \\
% \textit{Wageningen University and Research}\\
% Wageningen, The Netherlands\\
% tarek.alskaif@wur.nl}\\
% \and
% \IEEEauthorblockN{Declan O'Sullivan}
% \IEEEauthorblockA{\textit{ADAPT SFI Research Centre} \\
% \textit{Trinity College Dublin}\\
% Dublin, Ireland\\
% tarek.alskaif@wur.nl}
% \and
% \IEEEauthorblockN{Soumyabrata~Dev}
% \IEEEauthorblockA{\textit{ADAPT SFI Research Centre} \\
% \textit{University College Dublin (UCD)}\\
% Dublin, Ireland\\
% soumyabrata.dev@ucd.ie}
% }

\IEEEoverridecommandlockouts
% \IEEEpubid{\makebox[\columnwidth]{978-1-7281-1156-8/19/\$31.00~
% \copyright2019
% IEEE \hfill} \hspace{\columnsep}\makebox[\columnwidth]{ }} 

\maketitle
\thispagestyle{fancy}%

%---Content of Paper Abstract-----------------------

\begin{abstract}
% Today's smart energy network is driving toward the decentralization rapidly. 
In a decentralized household energy system consisting of various devices such as washing machines, heat pumps, and solar panels, understanding the electric energy consumption and production data at the granularity of the device helps end-users be closer to the system and further achieve the sustainability of energy use. However, many datasets in this area are isolated from other domains with records of only energy-related data. This may raise a loss of information (\textit{\textit{e.g.}} weather) that is relevant to the energy use of each device. A noticeable disadvantage is that many of those datasets have to be used in computational modeling approaches such as machine learning models, which are vulnerable to the data feed, to advance the understanding of energy consumption and production. Although such computational methods have achieved a high benchmark merely through a local view of datasets, the reusability cannot be firmly guaranteed when the information omission is taken into account. This paper addresses the data isolation problem in the smart energy systems area by exploring Semantic Web techniques on top of a household energy system. We propose an ontology modeling solution for the management of decentralized data at the resolution of a device in the system. As a result, the scope of the data concerning each device can be easily extended to be wider across the web and more information that may be of interest such as weather can be retrieved from the Web if the data are structured by the ontology.
\end{abstract}

\mbox{}
\nomenclature{\(HEC\)}{Household Energy Consumption}
\nomenclature{\(NOAA\)}{National Oceanic and Atmospheric Administration}
\nomenclature{\(CoSSMic\)}{Collaborating Smart Solar-Powered Microgrids}
\nomenclature{\(ICT\)}{Information and Communications Technology}
\nomenclature{\(CA\)}{Climate Analysis Ontology}
\nomenclature{\(RDF\)}{Resource Description Framework}
\nomenclature{\(UTC\)}{Coordinated Universal Time}
\nomenclature{\(PV\)}{Photovoltaics}

\printnomenclature[1in]

\section{Introduction}
\subsection{Motivation}
With the growing emergence of sensing technologies in the IoT area, today's HEC data can be generated at a decentralized level, such as in an electric vehicle, a heat pump, and the lighting. Due to the high variety and granularity of devices producing data, a new generation of smart energy systems is being steered towards decentralization and bears the great potential to facilitate a future of energy sustainability~\cite{van2020integrated}. The question for scientists is how to take advantage of the oncoming and vast ocean of data to achieve energy sustainability, namely energy cost reduction, emission reduction, and efficient use of energy.
However, analyzing household energy data, especially in a decentralized system, is becoming more challenging as a consequence of diverse smart devices interacting and composing a complex energy network~\cite{jain2021validating,jain2020clustering}. Moreover, researchers are likely to retrieve limited data from a specific data vendor, or collect data within their finite efforts, and then structure them for creating models~\cite{orlandi2019interlinking}. This may give rise to a considerable bias in the solution to a limited scale of smart energy problems, and hence the model's reusability is hindered by a foreseeable more complicated scenario where larger quantity, more types of data (\textit{e.g.} weather data), and the tendency of decentralization in smart energy area should be considered as the network evolves~\cite{ahmad2020review}. 
\subsection{Relevant Literature}
An emerging research perspective is to represent the data in an ontology model (defined in Section~\ref{sec:ontoshort}) such that the underlying relationships between data in an energy network can be articulated in human words, bringing intelligence to energy network analysis. Many studies has established ontologies for various smart energy application scenarios. SAREF4EE is an ontology developed by Daniele at al.~\cite{noauthor_undated-gh} for the optimization of energy demand and response. SEAS was introduced by Lefran\c{c}ois~\cite{Lefrancois2017-ag}, with the goal of enabling interoperability across smart energy sectors. Our study is inspired by prior ontologies for smart energy systems, but it focuses on the decentralized household energy systems as well as ease of inclusion of cross-domain impacts (\textit{e.g.} climate domain), which has received less attention from other researchers. To complete the climatic impacts modeling portion of our model, we also referred to Wu's~\cite{wu2021ontology} ontology CA and Janowicz's~\cite{Janowicz2019-hn} ontology SOSA in order to describe the climatic sensor data.

\subsection{Contributions and Organization}
In this work\footnote{\label{note3}In the spirit of reproducible research, all the source code is available at \url{https://github.com/futaoo/ontology-energysystems}}, we focus on these issues and propose to solve them with the creation of robust ontology models and the realization that the use of energy is also influenced by cross-domain factors, such as meteorology. 
% We show how it is possible to integrate these heterogeneous data sources, usually distributed separately on the Web, facilitating data collection, fusion, and pre-processing. We demonstrate how a well-defined schema could be leveraged for integrating these datasets and supporting HEC research purposes.

The novel contributions of this work are as follows:

\begin{itemize}
    \item Creating semantics for decentralized household energy sectors at a systematical level;
    \item Converting the household energy data to Linked Data~\cite{bizer2011linked} to improve the ease of including other domains' Linked Data throughout the Web;
    \item Connecting various meteorological variables with energy use in order to enhance the comprehension and analysis of the data.
\end{itemize}

The rest of this paper is organized as follows: in Section~\ref{sec:bkgd}, background information is provided, including the raw data description and some core semantic techniques used in this work. Section~\ref{sec:workflow} focuses on the whole workflow that transforms the local household consumption data into Linked Data, which grants the local data a web-wide accessibility. The advantage of Linked Data is discussed on the enrichment of NOAA weather data into the household energy data. Section~\ref{sec:analysishc} expands on the scope of local household energy data in a Linked Data platform by offering an example analysis of solar energy generation in accordance with temperature. Finally, in Section~\ref{sec:Conc}, we conclude this work and list some planned future works.

\section{Background}
\label{sec:bkgd}
\subsection{Data description}
\label{sec:datadesc}
\subsubsection{CoSSMic household energy data}
\label{sec:cossmic}
CoSSMic\footnote{\url{http://isc-konstanz.de/en/isc/institute/public-projects/completed-projects/eu/cossmic.html}} is a smart grid project funded by EU Framework Programme FP7 for Research and Innovation to investigate energy use optimization by developing intelligent microgrids for households in a German city--Konstanz~\cite{amato2017simulation}. The energy flow in the grid network is controlled by an autonomic ICT system where the energy consumption and the distributed energy generation are adapted in real time, in line with several criteria such as availability, price and weather conditions. This decentralized energy flow is achieved on the basis of coordinated load shifting, \textit{i.e.} power consumers and producers are able to negotiate with each other to optimize the energy exchange. The energy data during the investigation is available on the open power system data portal\footnote{\label{ftn:opsd}\url{https://data.open-power-system-data.org/household_data/}} and can be used for reanalysis purposes. However, the existing dataset only contains energy flow data and many other factors that are key to understand the energy exchange profile, such as weather data, are not available. The shortage of these factors in our consideration will impede the reanalysis purposes. In this paper, we attempt to address this problem to some extent by re-publishing the data on the Web via semantic techniques to obtain some other data sources for reference. This method will facilitate the usability of a local energy usage dataset.
\subsubsection{Linked climate data}
\label{sec:linkclimate}
\begin{figure}[ht]
\centering
\includegraphics[width=0.9\textwidth]{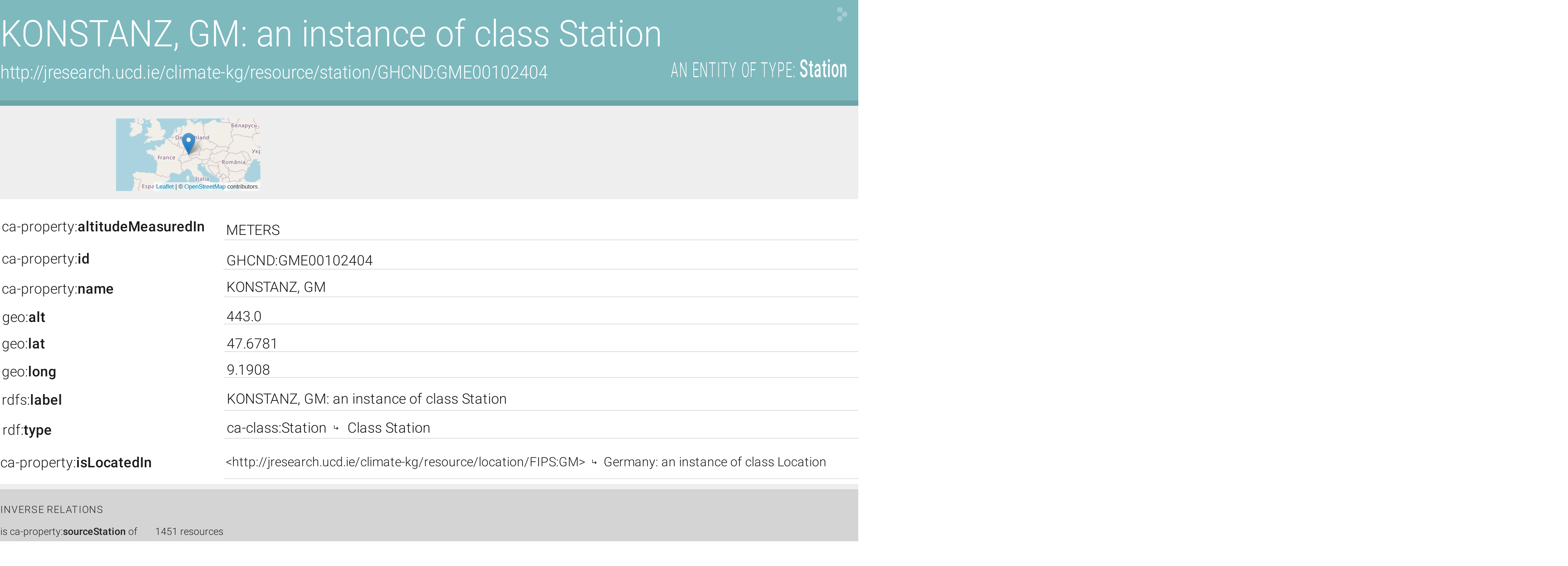} 
\caption{Visualization of the profile of a climate observation station in Konstanz.}
\label{fig:konstation}
\end{figure}

Linked climate data\footnote{\url{http://jresearch.ucd.ie/linkclimate/}} is a climate observation dataset that complies with Linked Data principles (details are given in Section~\ref{sec:shortsum})~\cite{wu2021ontology}. It provides NOAA Climate data\footnote{\url{https://www.ncdc.noaa.gov/cdo-web/}} kept in some Europe countries's (\textit{i.e.} covering Konstanz) stations through Linked Data platform. An example of a Linked Data view of a climate station in Konstanz can be seen on Fig.~\ref{fig:konstation}. The content of the climate observation station includes many types of meteorological observations such as temperature and precipitation. These data are open to the web and offer a flexible access to any published Linked Data in any domain if there is an appropriate ontology model that can establish the linkage to them.

\subsection{Short summary of semantics web techniques in use}
\label{sec:shortsum}
\subsubsection{Ontologies in short}
\label{sec:ontoshort}
Ontologies (or vocabularies in the simple term) unambiguously describe terms and relationships for particular fields of interest, according to the W3C definition\footnote{\url{https://www.w3.org/standards/semanticweb/ontology}}. As an example, in a wireless sensor network, ontology can be used to identify various types of sensors by simply calling them ``wind sensor'', ``temperature sensor'' and so on. Similarly, the related observations of meteorological variables can be linked to corresponding sensors by building links such as ``hasResults''. As a result of defining a collection of words, a semantic layer of operative human definable terms is built over the data.
% According to W3C's definition\footnote{\url{https://www.w3.org/standards/semanticweb/ontology}}: `ontologies' (or `vocabularies' in the simple term) unambiguously define concepts and relationships for specific fields of concern. Take wireless sensor network as an example, ontology can be made to classify different types of sensors by simply naming `wind sensor', `temperature sensor', etc. And the corresponding observations can be connected to by creating links named, for example, `hasResults'. Consequently with a set of terms defined, a semantic layer is formed consisting of operative human describable terms over the data. 
\subsubsection{Linked Data in short}
\label{sec:linkeddatashort}
Linked Data is a field of study that defines a collection of principles for how data is organized, linked, and published on the Internet~\cite{bizer2011linked}. It helps users to browse data by following data links from one data source to the next, allowing a fixed collection of data sources to be conveniently expanded with all of the other Linked Data sources available on the Internet~\cite{barbosa2021use}. One of the benefits of this activity is that, if necessary, federated queries (or reasoning) can be run on such a global large database on the Web to obtain resources (or knowledge) not available in fixed databases~\cite{8706177}.
% Linked Data is a research field and a set of principles that deal with how data is structured, interconnected, and published on the Web~\cite{bizer2011linked}. It allows people to navigate data through the data links from one data source to another, which enables a fixed set of data sources to be easily extended with all of the other Linked Data sources available on the Web~\cite{Radulovic2015}. One of the advantages brought by this behavior is such that --- if needed --- federated queries (or reasoning) can be performed on such global wide database on the Web to gain resources (or knowledge) beyond the fixed databases~\cite{8706177}. 
\section{Workflow towards Linked Data}
\label{sec:workflow}
\subsection{Data modeling}
\label{sec: onto}
\subsubsection{Ontology modeling for table headings}
\begin{figure}[ht]
\centering
\includegraphics[width=0.49\textwidth]{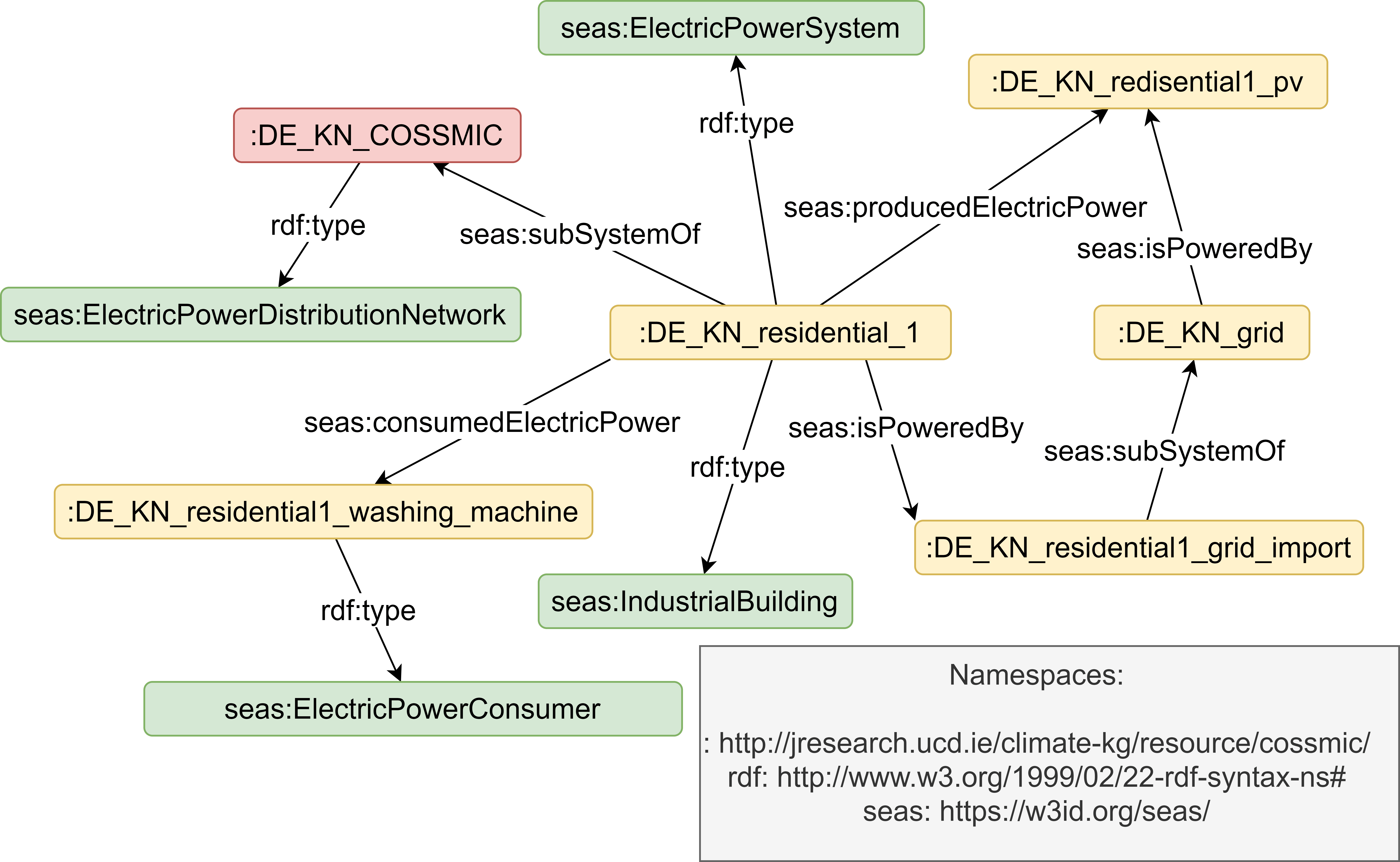} 
\caption{A graphical view of ontology model for some CoSSMic table headings where yellow nodes are derived from the headings, and classes are presented as green nodes; the red node ``:DE\_KN\_COSSMIC'' is modeled for the representation of the entire CoSSMic electric power distribution network.}
\label{fig:ontohead}
\end{figure}

Open power system data provides raw tabular CoSSMic data accompanied by the detailed documentation of the table headings. This paper first clarifies the meaning of the heading of each column in the documentation and identifies the relationships among individuals that can be extracted from the table headings. For instance, ``DE\_KN\_industrial1\_pv\_1'' is a heading with the annotation ``Total Photovoltaic energy generation in an industrial warehouse building in kWh''. Several individuals can be extracted: a country--German (``DE''), a city in the German--Konstanz (``KN''), an industrial building--industrial1 (``industrial1'') and a photovoltaic device--pv1 (``pv\_1''). If considering every individual associated with energy as a system in the energy network. Then the heading can be rephrased by the semantic statement that pv1 is a subsystem of industrial1 and industrial1 is a building located in a German city--Konstanz. Following on this rule, the set of table headings can be further expanded as relationships between the individuals in a heading and also relationships between individuals in different headings. To build a comprehensive ontology model for the CoSSMic data, we use SEAS knowledge model~\cite{seasd22} to describe the possible individuals and their relationships. Some main vocabularies used for clarifying classes and properties are outlined as follows\footnote{\textbf{Note}: ontology vocabularies in this paper are already associated with web addresses and comply with the form \{prefix\}:\{literal term\} where the the meaning of the prefix (name space) is given in the supplementary graphical representations of the vocabularies.}:
\begin{itemize}
    \item \textbf{seas:ElectricPowerDistributionNetwork}~~ (CLASS) denotes a network used to distribute the electric power;
    \item \textbf{seas:ElectricPowerTransmissionSystem}~~ (CLASS) denotes an electric power transmission system capable of transmitting electricity;
    \item \textbf{seas:isPoweredBy}~~ (PROPERTY) links a System to its powered system and the inverse vocabulary is ``seas:powers'';
    \item \textbf{seas:producedElectricPower}~~ (PROPERTY) denotes the produced electric power;
    \item \textbf{seas:consumedElectricPower}~~ (PROPERTY) denotes the consumed electric power;
    \item \textbf{seas:subSystemOf}~~ (PROPERTY) links a system to its super system.
\end{itemize}
A graphical representation using above vocabularies to model some extracted individuals from the table headings is displayed in Fig.~\ref{fig:ontohead}.

\subsubsection{Ontology modeling for data instances}
To turn the tabular data into the RDF data for the further storage as Linked Data in the following steps, this step continues the use of seas ontology to represent concrete data instances and the relationships between them and their corresponding headings (data fields) in the CoSSMic dataset. This paper assumes each record in regarding to energy consumption and production from every device as an ``Evaluation'' in the seas ontology and uses the vocabularies ``seas:consumedElectricPower'' and ``seas:producedElectricPower'' mentioned above to distinguish the data between the energy consumption and production.  The main vocabularies for modeling data entries are listed below:
\begin{itemize}
    \item \textbf{seas:ElectricPowerEvaluation}~~ (CLASS) denotes evaluations for electric power properties;
    \item \textbf{seas:evaluation}~~ (PROPERTY) links a valuable entity to one of its evaluations.;
    \item \textbf{seas:evaluatedValue}~~ (PROPERTY) links an evaluation to the literal (numeric value in this paper);
\end{itemize}

\begin{figure}[ht]
\centering
\includegraphics[width=0.49\textwidth]{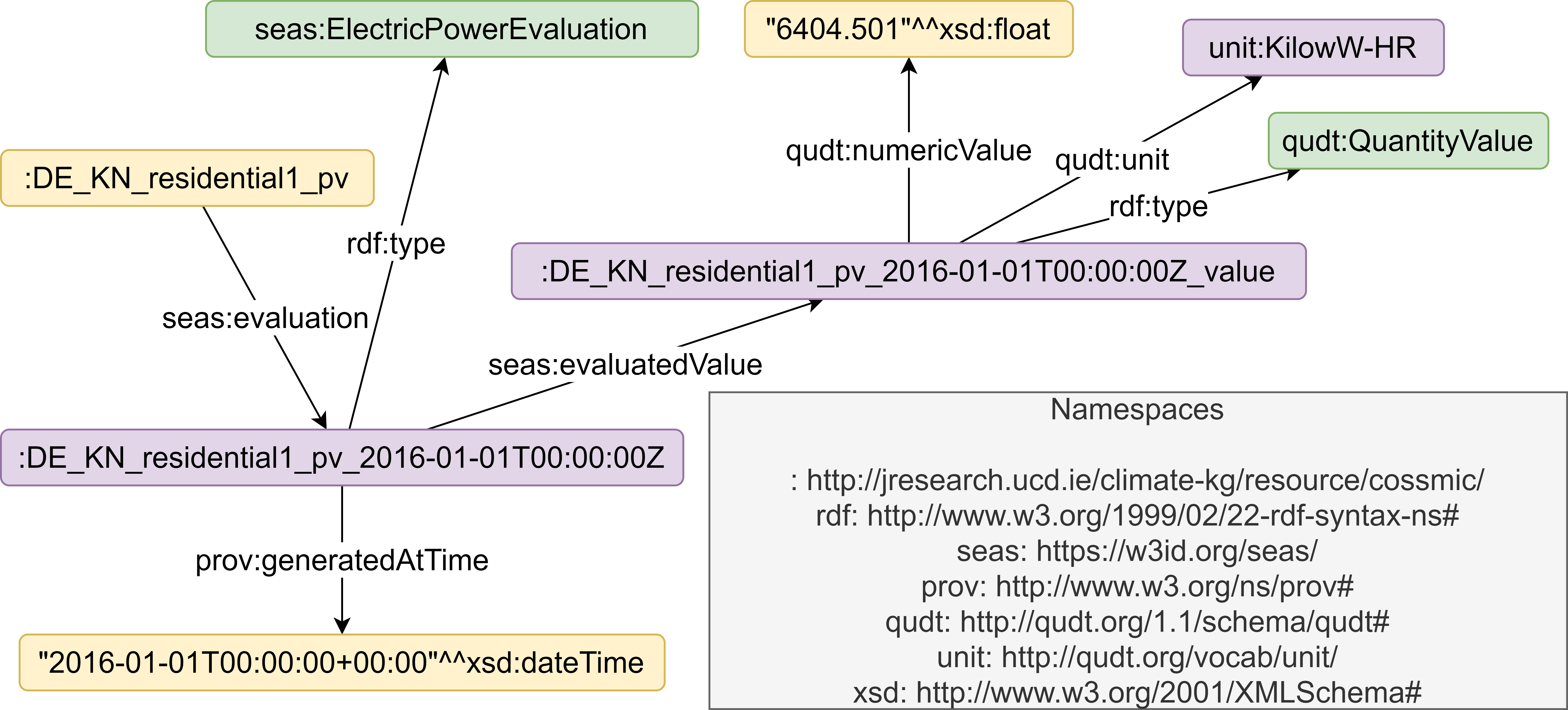} 
\caption{A graphical view of ontology model for CoSSMic data entries where the node ``DE\_KN\_residential1\_pv'' is identical to the one in Fig.~\ref{fig:ontohead} ,and yellow nodes are derived from CoSSMic dataset; Coordinated Universal Time (UTC) is used for the time.}
\label{fig:ontodata}
\end{figure}

\subsection{Uplifting NOAA data}
\label{sec:NOAA}
\begin{figure}[ht]
\centering
\includegraphics[width=0.49\textwidth]{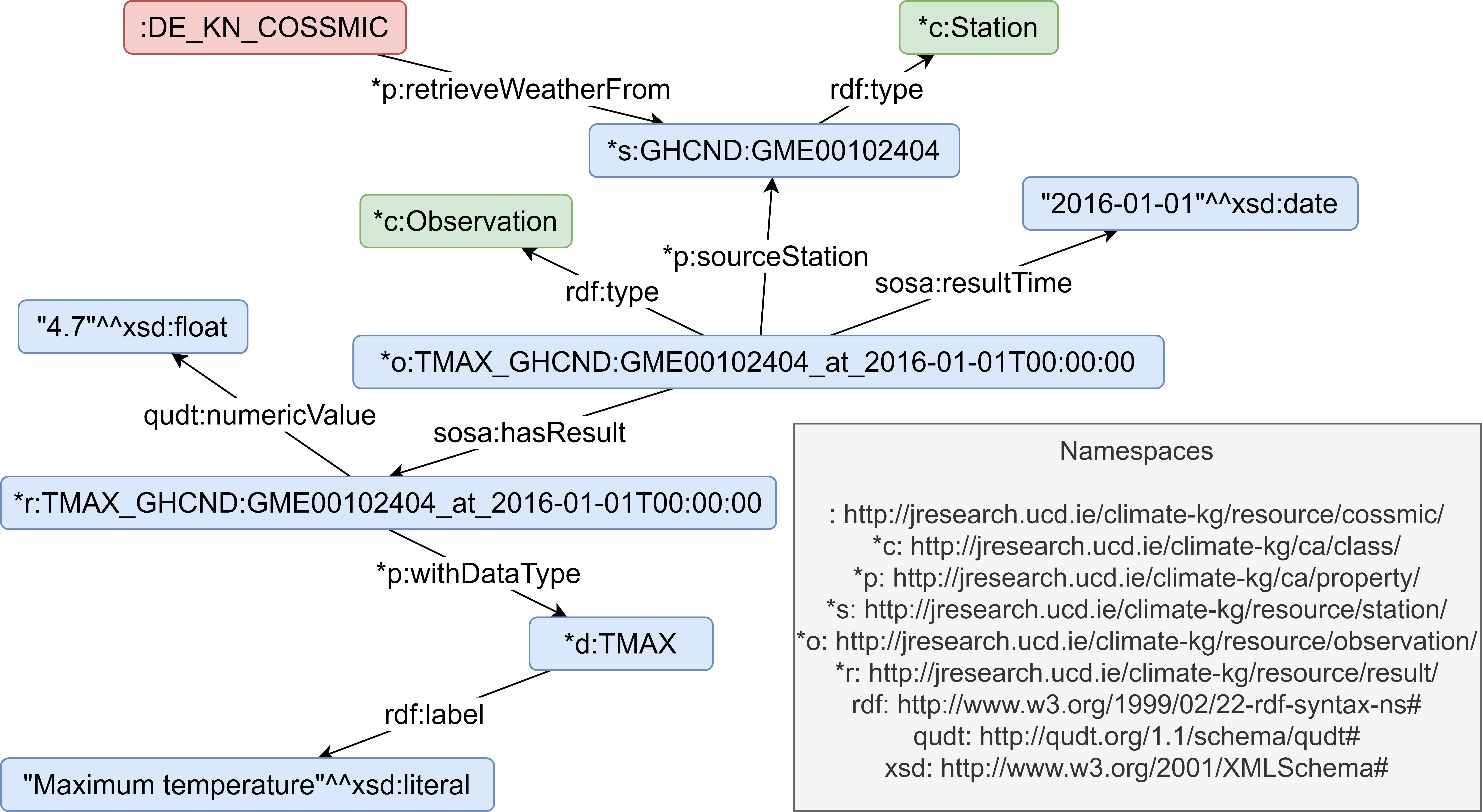}  
\caption{A schematic diagram represents integration of NOAA climate data at semantic level; the red node ``:DE\_KN\_COSSMIC'' is identical to the one in Fig.~\ref{fig:ontohead}; The blue nodes represent the modeled individuals of linked climate data; the green nodes clarify the classes; UTC is used for time recording.}
\label{fig:hec}
\end{figure}

During the course of steps in Section~\ref{sec: onto}, an energy distribution network is constructed with semantic techniques. Thanks to the ontology model, it is easier to incorporate other Linked Data sources available on the Web by defining proper ontology terms. This will bring more information beneficial to understand the household energy consumption and generation. We create a new term ``retrieveWeatherFrom'' in the CA ontology~\cite{wu2021ontology} (the ontology defined for linked climate data in Section~\ref{sec:linkclimate}) to achieve the semantic integration of meteorological data into the energy network. The meteorological data used for modeling purposes are from our published linked climate data as mentioned in Section~\ref{sec:linkclimate}. To understand the structure of the linked climate data, we list some vocabularies used in CA ontology and a graphical view (Fig.~\ref{fig:hec}) of one temperature observation from a Konstanz weather station:
\begin{itemize}
    \item \textbf{*c:Station}~~a CLASS denotes a station that observes some feature of interest such as precipitation, temperature, etc.;
    \item \textbf{*c:Observation}~~a CLASS denotes an observation of some feature of interest;
    \item \textbf{*p:sourceStation}~~a PROPERTY links an observation to the station that it belongs;
    \item \textbf{*p:withDataType}~~a PROPERTY links the data to its data type;
    \item \textbf{*p:retrieveWeatherFrom}~~a PROPERTY links an individual to the individual that provides the weather information;
    \item \textbf{sosa:hasResult}~~a PROPERTY links an observation to its result.
    \item \textbf{sosa:resultTime}~~a PROPERTY links an observation to the time when the observation is generated.
    
\end{itemize}

\subsection{Conversion to Linked Data}
\label{linkeddata}
In this phase, we convert the CoSSMic data from tabular to graph format, i.e. to RDF. The CoSSMic RDF data is structured by the ontology model devised previously. We create a Python script using RDFLib to map and transform all tabular data entries to RDF data and, in the meanwhile, store the resulting RDF into the Fuseki\footnote{\url{https://jena.apache.org/documentation/fuseki2/}} Linked Data platform. There are several noticeable advantages of publishing CoSSMic data in RDF. Any potential community of interest is capable of accessing the data over HTTP. Other data sources can favorably integrate the CoSSMic data through Linked Data techniques and manipulate the data at the individual level at most, instead of downloading a bulk of CSV data. It also provides the CoSSMic data with the possibility to integrate more information other than NOAA climate data, for example, to obtain air pollution, green house gas emissions, and more meteorological data in other Linked Data platforms. In the Section~\ref{sec:analysishc}, we will demonstrate how to perform queries on the Linked Data platform and give an example of analysis on climate data and CoSSMic data.
\section{Analysis of household energy data and climate in our Linked Data platform}
\label{sec:analysishc}
\subsection{Semantic queries on the combination of HEC and climate data}
\label{sec:query}
The published endpoint offers some options to acquire the linked HEC data and climate data. Users could either perform queries directly on endpoint interface\footnote{\url{http://jresearch.ucd.ie/kg/dataset.html?tab=query&ds=/climate}} or embed the queries as HTTP requests in their code. The latter one is supported by some offered HTTP APIs in the endpoint server and can be more efficient if more complicated evaluations to the queries are involved (\textit{i.e.} consuming computing resources at the client side). The queries should be written in standard SPARQL 1.1 language to define a graph pattern for getting the solutions of the queries\cite{world2013sparql,hogan2020knowledge}. An example query to obtain data regarding the time series of solar energy generated by a PV device and its corresponding weather alignments is given in Listing 1.

% \begin{listing}[ht]
% \includegraphics[width=0.5\textwidth]{listing.png} 
% \caption{A sample query that retrieves a time series of daily solar energy production with alignments of daily maximum temperature from the published SPARQL endpoint.}
% \label{lst:sqarql}
% \end{listing}

\begin{figure}[htb]
\centering
\includegraphics[width=0.5\textwidth]{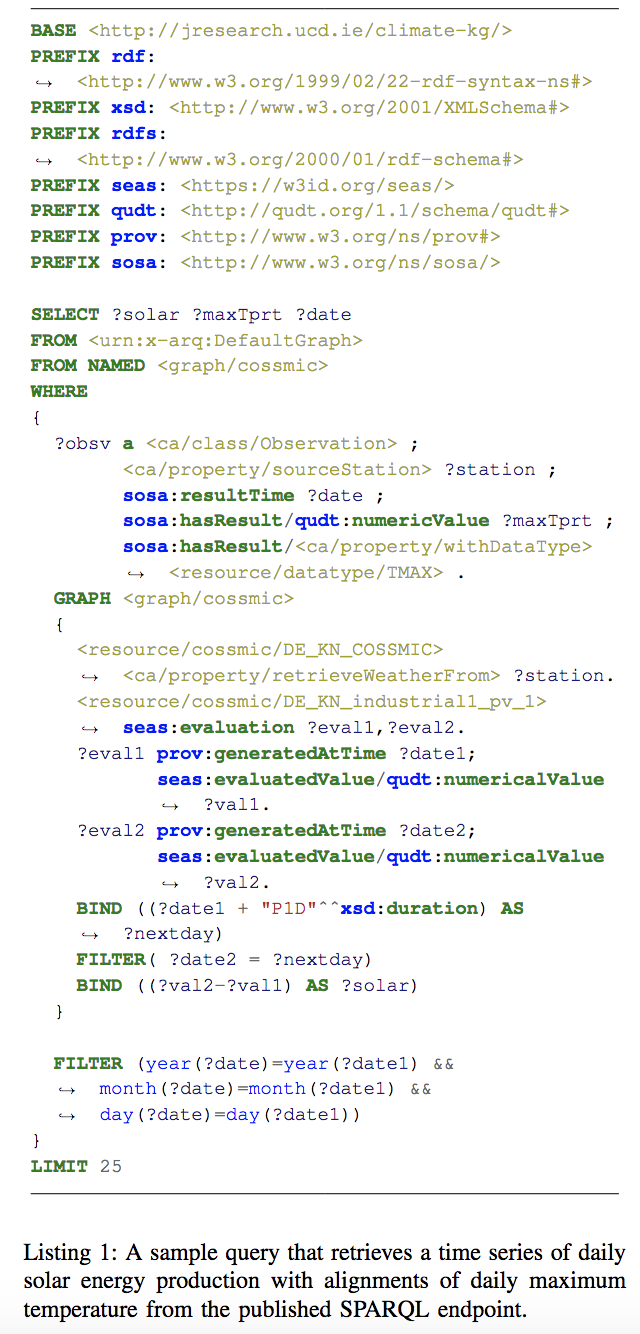}
\end{figure}

\subsection{An example of solar energy generation against temperature}
\label{sec:solar}
\begin{figure}[ht]
\centering
\includegraphics[width=0.5\textwidth]{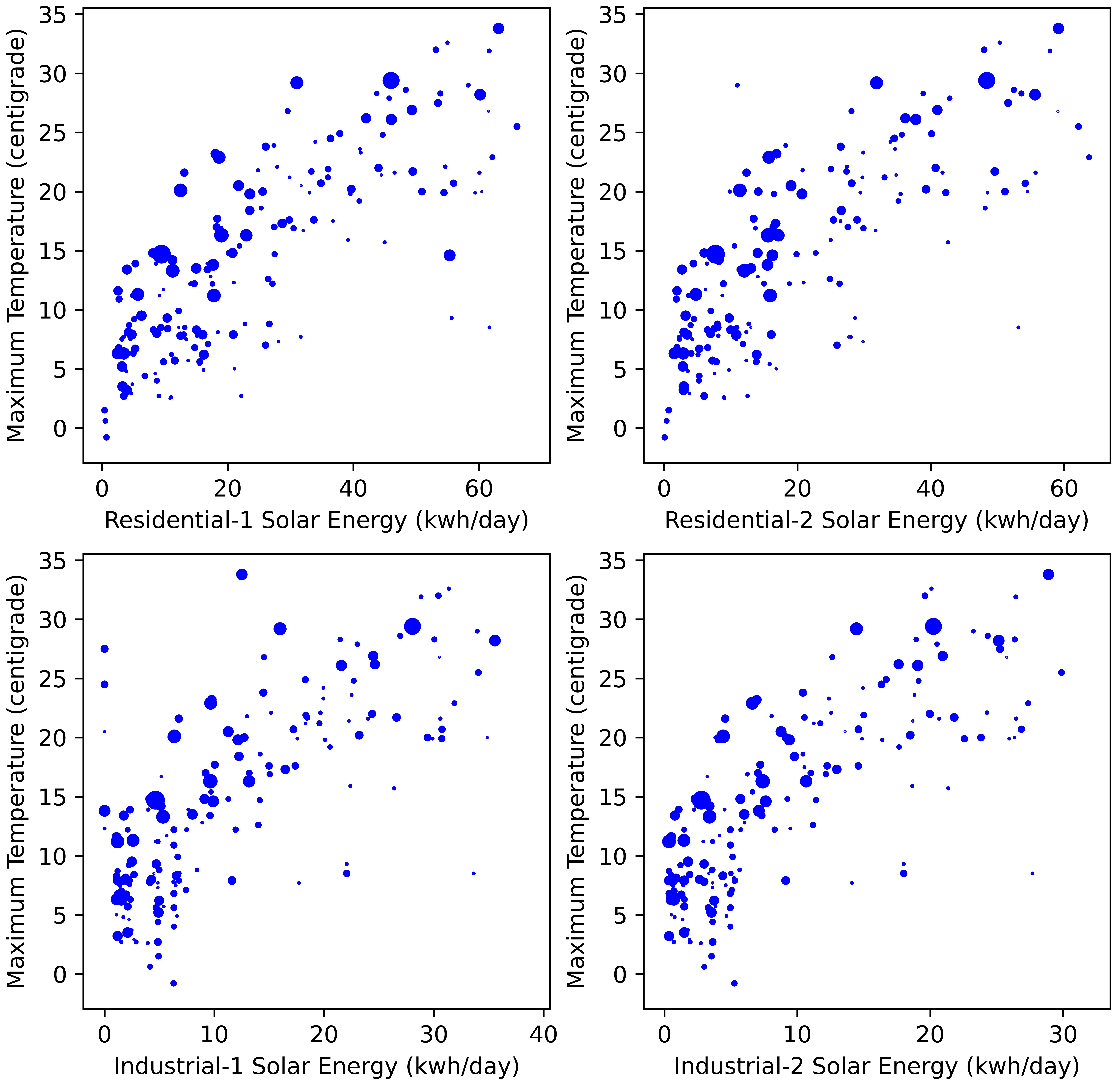} 
\caption{Daily solar energy generation against daily maximum temperature with precipitation amount in the form of scatter size.}
\label{fig:solar}
\end{figure}

Establishing a linkage from HEC to climate data offers a new view to the HEC. The climate information integration can be used to perform a more comprehensive analysis regarding the weather influence on energy consumption and generation. As an example case, we select four PV devices (\textit{i.e.} 2 residential buildings and 2 industrial buildings) previously located in the CoSSMic dataset for the solar energy generation data during 2016 with the addition of daily maximum temperature and daily precipitation information from the published endpoint to perform correlation analysis. Daily solar energy vs. daily maximum temperature for these four devices (precipitation amount is reflected by the size of the scatter) are shown in Fig~\ref{fig:solar}.

The figure illustrates a strong positive correlation between the solar energy and max temperature. To be more accurate, the calculated Pearson correlation coefficients~\cite{schober2018correlation} between meteorological variables (\textit{i.e.} maximum temperature and precipitation) and solar energy generation are shown in Table~\ref{table:pearson}. The correlation coefficients associated with solar energy generation are approximately 0.8 on average to maximum temperature, however roughly -0.15 on average to the precipitation (\textit{i.e.} precipitation has a negative linear influence on the solar energy). This result is also helpful when performing the feature selection for modeling tasks and the maximum temperature could be included as one of the features to enhance the modeling tasks~\cite{alskaif2020systematic, liu2020daily, manandhar2018systematic}.

\begin{table}
\setlength\tabcolsep{0pt} % make LaTeX figure out the intercolumn whitespace amount
\caption{Pearson Correlation Coefficients Between Meteorological Variables and Household Solar Energy Generation.} \label{table:pearson} 
\sisetup{round-mode=places,round-precision=2} 
\begin{tabular*}{0.48\textwidth}{@{\extracolsep{\fill}} l *{14}{S[table-format=-1.2]} }
\toprule
 & {Residential-1} & {Residential-2} & {Industrial-1} & {Industrial-2} \\
\midrule
TMAX & 0.792388	& 0.782947 & 0.779747 & 0.802174 \\
PRCP & -0.190667 & -0.164119 & -0.140355 & -0.124258 \\
\bottomrule
\end{tabular*}
\end{table}

\subsection{Discussion}
In this section, we discuss the role of semantic techniques in enhancing a local dataset with more details from the following aspects:

\paragraph{Adding new datasets}
\label{para:addnew}
To add a new dataset into a SPARQL endpoint as part of the Web of data (or Linked Data), authors need to define an ontology to structure the dataset in a similar way to the workflow proposed in this paper. Users who are interested in using the data published by others also need to know the ontology of the data so that being able to compose graph patterns with SPARQL language to select the data of concern. 

\paragraph{Interlinking diverse data}
\label{para:interlink}
In many cases like cross-domain analysis, heterogeneous data sources need to be fused into a dataset. Semantified heterogeneous data can be efficiently integrated with other specialized Linked Data fields at ontology level. However, this requires a certain degree of expert domain knowledge in the schema of all relevant datasets and a number of appropriate ontology vocabularies to perform the semantic enrichment\cite{patricio2020web,musyaffa2020iota}. 

\paragraph{Using more data to understand smart energy} In this work, we demonstrate one data source (NOAA meteological variables) which is beneficial to improve the understanding of a smart energy system in Konstanz. Many other fragments of data also have the potential to be ingested to this end, such as solar radiation data\footnote{\url{https://www.dwd.de/EN/ourservices/solarenergy/maps_globalradiation_average.html}}, other weather data (\textit{e.g.} meteoblue\footnote{\url{https://www.meteoblue.com/en/weather/archive/export/konstanz_germany_2885679}}), places of interest (\textit{e.g.} OpenStreetMap\footnote{\url{https://www.openstreetmap.org/}}). Assimilating these datasets should comply with the principles mentioned above (Section~\ref{para:addnew} and Section~\ref{para:interlink}).

\paragraph{Federated queries across SPARQL endpoints} The data published as Linked Data has not to be in one SPARQL endpoint. The implementation of standard or extended SPARQL endpoints supports federated queries with other endpoints, \textit{i.e.}, data distributed into various remote endpoints are able to be directly used as sources and are retrievable in any endpoint by the SPARQL queries\cite{grall2020collaborative}. An assemble of open Linked data sources is gathered by Linked Open Data\footnote{\url{https://lod-cloud.net/}} where users can acquire data serviceable to the smart energy data processing.

\section{Conclusion and Future Work}
\label{sec:Conc}
In this paper, we devise a workflow to transform a local decentralized HEC dataset (CoSSMic) into a Linked dataset and republish them across the web. This work shows several detailed steps to complete this transformation and as a result, the scope of the local dataset is expanded and broader views of the weather information have been included to the dataset by semantic techniques. The final analysis conducted on solar energy production with the temperature and precipitation at device level also demonstrates the weather information from other sources is helpful for modeling tasks. In the future, we aim to work on more semantic techniques such as Geosparql and temporal RDF to create a powerful interoperable Linked Data integration platform to be able to benefit many CoSSMic HEC likewise local datasets so that it can be used as a way to enhance the understanding of the smart energy data.

% Generated by IEEEtran.bst, version: 1.14 (2015/08/26)

\balance 

%--------------------------

\end{document}